\documentstyle[pre,aps,preprint]{revtex}
\tightenlines
\begin{document}
\title{Dynamical quenching and annealing in self-organization multiagent models}
\author{E. Burgos$^1$\thanks{e-mail burgos@cnea.gov.ar}, Horacio
Ceva$^1$\thanks{ e-mail ceva@cnea.gov.ar}, R.P.J.
Perazzo${^{2,3}}$\thanks{perazzo@df.uba.ar}}
\address{$^1$Departamento de F{\'{\i}}sica, Comisi{\'o}n Nacional de
Energ{\'\i }a At{\'o}mica,\\
Avda. del Libertador 8250,1429 Buenos Aires, Argentina \\
$^2$Departamento de F{\'{\i}}sica FCEN,Universidad de Buenos Aires,\\
Ciudad Universitaria - Pabell{\'o}n 1, 1428 Buenos Aires, Argentina\\
$^3$Centro de Estudios Avanzados, Universidad de Buenos Aires,\\
Uriburu 950, 1114 Buenos Aires, Argentina}
\date{21 December 2000}
\maketitle

\begin{abstract}
We study the dynamics of a generalized Minority Game (GMG) and of the Bar
Attendance Model (BAM) in which a number of agents self-organize to match
an attendance that is fixed externally as a control parameter.  We compare
the usual dynamics used for the Minority Game with one for the BAM that
makes a better use of the available information.  We study the asymptotic
states reached in both frameworks.  We show that states that can be
assimilated to either thermodynamic equilibrium or quenched configurations
can appear in both models, but with different settings.  We discuss the
relevance of the parameter $G$ that measures the value of the prize for
winning in units of the fine for losing.  We also provide an annealing
protocol by which the quenched configurations of the GMG can progressively
be modified to reach an asymptotic equlibrium state that coincides with
the one obtained with the BAM.\\{PACS Numbers:  02.50.Le,05.40.-a,
64.60.Ak, 89.90.+n\\}
\end{abstract}



\section{Introduction}

In recent times considerable attention has been given to the description of
self-organization processes in multiagent models. These are geared to
explain how (and which) individual decisions give rise to cooperative
effects that show up in the behavior of the system as a whole. Such models
have a wide number of interesting applications in a variety of situations
that range from routing of messages in an information network to the way in
which equilibrium is (dynamically) achieved by an ensemble of economic
agents.

The main question posed in this line of analysis is how a system composed of
multiple agents acting in a decentralized fashion is able to reach a state
of ``macroscopic order'' in which individual decisions appear to be
coordinated with one another. A well know example of this line of research
is the Bar Attendance Model (BAM) proposed by B. Arthur in Ref~\cite{Arthur2}.
 This model has been developed to illustrate a case of coordination through
a process that can be assimilated to inductive learning.

In the BAM a (large) number of individuals have to decide whether or not to
go to a bar at a certain date. They share the common perception that
attendance is not desirable when the number of people who are present
exceeds a critical value, which is smaller than the total population of
clients. Otherwise they derive utility from going to the bar.

A dynamics is proposed by which each agent individually updates her criteria
to decide the attendance. These are based upon public information, that
concerns the total attendance in the previous days. Agents use inductive
reasoning to predict the number of customers in the following day.

It is simple to realize that the model is self-deceiving: if everybody
decides to attend, crowding is produced and everybody loses. Everybody will
then refrain to go the next day situation in which, again, everybody loses
because many could go without crowding. In an iterated version of the model
agents plan their attendance for the next few days. In this case it is more
evident that the only stable solution is a situation in which a diversity of
individual strategies compensate with each other saturating the capacity of
the bar every day. The system self-organizes in a (dynamic) Nash equilibrium
in which every agent optimizes their respective situation given what has
been done by the rest of the players. In this case any player that changes
unilaterally her rules ends up loosing. Such equilibrium is also Pareto
optimal because no agent can improve her situation without harming that of
some other player. If adaptation is stocastic, the attendance equilibrates
at the accepted value with random fluctuations.

An alternative version of the BAM called the Minority Game (MG) has recently
received much attention \cite{Challet},\cite{Savit},\cite{Cavagna},\cite
{Johnson}. The main feature of this framework is that agents derive utility
whenever their choice coincides with that of the minority. Self deceiving is
also evident within this framework. The MG has been presented as a
simplified version of the BAM. In the present paper we concentrate in the
approach of the MG introduced in ref.\cite{Johnson} and make a detailed
comparison of the MG and the BAM. We will see that the MG, besides being a
particular case of the BAM (namely one in which the optimal attendance has
been fixed at half of the total number of customers) it makes a different
use of the same information \cite{bcp}.

The treatments of the BAM have traditionally~\cite{Arthur2},\cite{nosotros}
been based upon genetic algorithms\cite{Goldberg}. Each agent is assumed to
keep a set of strings that encode her possible attendance schedules for the
following days. After that period each agent computes the outcome of all her
possible plans and chooses the best to use it during the next period. The
attendance schedules are considered as the ``genomes'' of a genetic
algorithm and are updated by selection, random mutations and crossover. The
BAM can be regarded as an iterated version of the MG, i.e. one in which
players have to make their choices for several rounds of the same game.
There is however one important difference. The BAM only involves (the
iteration of) pure strategies where players either go or not go to the bar.
The players in the MG have instead {\em mixed} strategies, i.e. they make
their choice of going or not going {\em with a given probability}. The
iterated pure strategies in the BAM undergo an adaptive process. Within the
MG mixed strategies also change. Each agent randomly attends the bar (with a
given probability) and collects points or pays fines depending that the
attendance is below or above the acceptable level. The mixed strategy
(i.e.her attendance probability) is changed when her account falls below a
given threshold. Both the BAM and the probabilistic approach of ref.\cite
{Johnson} can be regarded as mean field descriptions because only the
aggregate effect of the whole ensemble of agents is relevant for both
adaptive processes. On the other hand, the mixed strategy played by the
agents in the MG can be assimilated to a time averaged version of the
iterated attendace schedules of the players in the BAM.

In order to render possible a comparison between the MG and the BAM two
elements have to be considered. On the one hand a generalized version of the
MG (the GMG) has to be developed to describe situations traditionally
considered within the BAM, namely cases in which the winning choice is to
meet the one taken by an {\em arbitrary} fraction of the population, not
just the minority. There are no systematic studies of this GMG. On the other
hand a probabilistic (mixed strategy) mean field approach of the BAM has to
be developed. There are a number of possibilities to define the mixed
strategies for these two models. In the present paper we consider for the
GMG an obvious extension of the adaptive updating dynamics that reproduces
the well known selforganizing features of the MG. For the BAM we propose
instead a mixed strategy that reproduces the results obtained with the usual
treatments of this model that use iterated, pure strategies. It turns out
that both dynamics are different. It happens that the one associated to the
BAM makes a better use of the available information than in the GMG.
Although in both frameworks - GMG and BAM - all agents have access to the
{\em same} information, with the proposed BAM rules they are able to keep
better track of the causes for winning or loosing thus making a more
effective use of it than in the GMG. These seamingly minor changes produce
widely different cooperative behaviors. The GMG leads in general to quenched
states that are far from equilibrium. The BAM evolution leads instead to an
asymptotic configuration that can be thought of as a thermodynamic
equilibrium: it has lost memory of the initial conditions, the individual
strategies continue to be updated at all times and yet the density
distribution of attendance strategies remains stationary. In the following
sections we discuss how both models yield similar results in a rather
reduced portion of the space of parameters but turn out to be largely
different in all the rest. We show that an essential parameter that governs
the self-organized configuration of the system is determined by the rewards
and punishments that are contained in the scoring of successes and failures
of the agents. The control parameter is the value $G$ of the prize for
winning relative to the fine for losing. Ordinarily the effect of this
parameter has been disregarded and the only case studied in the literature
is $G=1$. Quenching and equilibrium regions change significantly for $G\not
=1$.

In the next Section we describe and characterize the quenched state in the
GMG and discuss the phase diagram that is found if $G$ is left as a free
parameter. In Section 3 we compare this to the BAM. In Sect. 4 we provide an
analytic discussion to support the results presented in the previous
sections. In Sect. 5 we present an annealing protocol that makes use of the
fact that quenching is a consequence of the memory stored in the system
encoded by the points gained by the agents. The protocol allows to reach
equilibrium using the MG dynamics by modifying progressively the quenched
state removing the memory stored in the system.

\section{The GMG model}
\label{sec:GMG}

We consider an $N$-agent system in which each has an mixed attendance
strategy expressed by a probability $p_i$ ($i=1,2,\dots ,N$) to go to the
bar. We take $n$ to be the acceptable level of attendance without crowding.
The prescription to adapt the individual strategies is the following. If the
attendance $A$ turns out to be $A\leq n$, and the agent chooses to go to the
bar, she gains $G$ points. Otherwise, pays a fine of one point. The value $G$
is left as a free parameter .

The balance of gains and losses keeps a record of successes and failures of
each agent. If the latter are more frequent than the former, her account
ends up by falling below some fixed threshold (that we take to be 0). In
this case her account is reset to zero and her strategy $p_i$ is updated.
This is done by changing the value of $p_i$ by one chosen at random within
the interval $(p_i-\delta p,p_i+\delta p)$. The witdth of this interval (we
take $\delta p=.05$) is left unchanged during the whole process. Reflective
boundary conditions for $p_i$ are used whenever necessary.

All $N$ agents have initially 0 points and their initial attendance
strategies are chosen with some criterion. In the present section we
restrict ourselves to consider the case in which all $p$'s are initially
chosen at random from the interval $[0,1]$ with uniform probability.

The state of the many agent system is well described by the density
distribution $P(p)$ that gives the fraction of agents that have a strategy
within the interval $(p,p+dp)$. In order to calculate this we generate an
ensemble of $H$ histories of the $N$-agent system by evolving it during a
fixed number of iterations and starting from statistically equivalent
initial conditions. The density $P(p)$ is obtained by normalizing the
frequency of occurrence of a given strategy in the whole set of the $H$
histories. The number of iterations is fixed by checking that a stationary
configuration has been reached.

We also compute other parameters that describe the state of the system. One
is the attendance $A(t)$ averaged over the $H$ histories as a function of
the iteration number $t$; in the following we quote $M(t)\equiv A(t)/N$, to
facilitate the comparison for different values of $N$. A second important
parameter is the fraction of agents $C_s(t)$ that change their individual
strategies as a function of the iteration number. This fraction provides a
direct measure of the stability of the state that has been reached. We will
use $P_{\infty}$ to refer to the asymptotic distribution, $t\rightarrow
\infty ,$ {\em whenever }the system reaches an stationary state.

A well known result obtained within the framework of the pure MG is a strong
polarization of the population. A distribution $P_\infty ^{MG}(p)$ is
reached in which essentially half of the agents always go to the bar while
the other half never go . As a consequence $M(t)\rightarrow 1/2$ except for
random fluctuations due to the updating of individual strategies that take
place at all times \cite{Johnson},\cite{bc}.

The evolution of a system with the GMG rules and $n\not =N/2$ has been
studied by Johnson \cite{Johnson2} and by us~\cite{bcp}. If the reduced
acceptable level of attendance $\mu \equiv n/N$ is close enough to 1/2 an
equilibrium situation is reached that resembles the MG results (see Fig.\ref
{ejemplos1}a)). Equilibrium is shown by the fact that $M\rightarrow \mu $
and $C_s(t)$ stabilizes at a constant, non zero value after a transient. In
Fig.~\ref{ejemplos1}b) we show typical plots of $M(t)$ and $C_s(t)$ for this
situation. On the contrary, if $\mu $ is larger than a critical value $\mu
_{cr}$ (see below) a situation is reached in which no further updating of
the strategies takes place ({\em 'quenched'} configuration). In Fig.~\ref
{ejemplos2} we show an example of the distribution $P_q(p)$ that is obtained
by setting $\mu >\mu _{cr}>1/2$.

The distribution $P_q(p)$ that is obtained strongly differs from $P_\infty
^{GMG}(p)$. The fact that $C_s(t)$ tends exponentially to 0 (see Fig.\ref
{ejemplos2}b)) indicates that a situation has been found that is completely
different than the one commented before. This happens because the largest
fraction of the population has gained many points and even though their
accounts may fluctuate, they are not expected to fall below the threshold
required to update their individual strategies. The consequence of this is
that $M$ stabilizes at a value that may be far from $\mu $. While the
situation for $\mu \sim 1/2$ can be thought as some kind of thermodynamic
equilibrium, this case should instead be associated to a quenched
configuration in which the state variables are rapidly fixed at values that
are close to the initial conditions.

We have performed extensive numerical simulations to investigate this
situation using values of $\mu $ above and below $1/2$ and several values of
$G$. As expected an overall symmetry is found around $\mu =1/2$. This is
because matching an attendance level greater than $1/2$ is equivalent as
matching a symmetric {\em absence} level equally greater than $1/2$. As will
shortly become evident symmetry is not strict. We discuss this point in
Section 5.

Several features that are displayed in the quenched distribution shown in
Fig.~\ref{ejemplos2} are worth to be remarked. In the first place it is
found that $P_q(p)=0$ for $0<p<1/2$ and $\mu \gg 1/2$. This can easily be
understood. When $\mu \gg 1/2$ the agents that have $p<1/2$ lose more
frequently that they win and therefore update their attendance strategies
more frequently increasing the value of $p$ whenever possible. Once their
individual strategies are greater that $1/2$ the inverse happens, agents
seldom lose points and therefore are not forced to update their individual
strategies. This also justifies that the width of the hump near $p=1/2$
turns out to be only a function of $\delta p$.

The above arguments indicate that the shape of $P_q(p)$ strongly depends
upon the initial conditions. This is the reason why we prefer to call this
effect ``quenching'' rather that ``freezing''. The latter suggests a new
ordering that bears no relationship with the initial conditions, while the
former indicates that a dynamics has been imposed that suddenly reaches a
fixed point leaving the internal parameters of the system at values that are
close to the ones chosen with the initial conditions and that may be far
from equilibrium.

In Fig.~\ref{Concurrencia} we illustrate some cases in which quenching takes
place. We plot the asymptotic value of the reduced attendance $M$ reached by
the system as a function of $\mu $, displaying only the sector with $\mu
>1/2 $, for several values of $G.$

In Fig.(\ref{Concurrencia}b) we also show a plot of the dispersion $\sigma
_M $ of the values of $M$ obtained in all the $H$ histories of the
ensemble ($\sigma_M^2=\sum_h^H(A_{\infty}(h)-<A_{\infty}>_H)^2)/NH $).
This parameter plays the r{\^o}le of a susceptibility. It displays a peak
located at the transition region and is essentially constant outside it. The
finite width of the peak as well as the irregularities of $M$ in the
transition region are to be attributed to the finiteness of the system.

It is possible to associate a critical value $\mu _{cr}$ with the change
between both regimes. In the numerical experiments, the large fluctuations
of $M$ near $\mu _{cr}$ are due to the fact that some of the histories that
make up the ensemble happen to quench while others, purely due to random
fluctuations, reach instead equilibrium. As $\mu $ increases, the proportion
between both types of histories gradually changes leading to the situation
in which all possible evolutions lead to quenched states. The peaked
behavior of $\sigma _M$ at $\mu =\mu _{cr}$ supports the picture that a sort
of critical phenomenon actually takes place in the system for a precise
value of $\mu $.

There are several criteria to define $\mu _{cr}$. We have chosen to define
it by the minimum value of $\mu $ for which the dynamics leads the system to
a situation in which the whole population has accumulated a fixed gain, that
we (arbitrarily) chose to be not less than 20 points. In Fig.(\ref
{Concurrencia}b) we show $\varphi _{20},$ the fraction of the population
that has accumulated at least 20 points as a function of $\mu $. It is clear
that if this happens any further changes in the distribution $P(p)$ or in
the attendance are extremly unlikely to occur. Another possibility is to
choose the value of $\mu $ that corresponds to the intersection of the lines
that are associated to both regimes or the maximum of $\sigma _M$. Again the
difference between these deffinitions are to be attributed to the finitness
of the systems that we have studied.

The value of $\mu _{cr}$ depends both upon $G$ and $N$. Changes in $G$ have
not been explored previously and cause important changes in the region in
which quenching occurs as well as in the shape of the distribution $P_\infty
^{GMG}(p)$. The effects of the value of $G$ in $\mu _{cr}$ can be seen in
Fig.~\ref{Concurrencia}a). The general phase diagram (see Fig.\ref{Fases})
shows two regions that correspond to quenched and unquenched systems. The
border between both regions is given by the function $\mu _{cr}(G)$. Finite
size effects are also displayed in this figure through the values of $\mu
_{cr}(G)$ for two values of $N$. The results displayed in Fig.~\ref{Fases}
allows to conjecture that for an infinite system with $G=1$ and $\mu =1/2$
the usual settings of the Minority Game lead to an unstable quenched phase.

The changes in the distribution $P_\infty ^{GMG}(p)$ that are due to changes
in $\mu $ keeping $G$ fixed are shown in figs \ref{ejemplos1} and \ref
{ejemplos2} and \ref{baldes}a). An increase in $\mu $ produces an
increasingly asymmetric distribution. This hold up to the point in which
$\mu >\mu _{crit}$ and quenching occurs giving rise to a humped
distribution
$P_q^{GMG}(p)$. The changes introduced by different values of $G$ are
shown
in Fig.~\ref{baldes}b). The overall effect of this parameter is to increase
the polarization, preserving the level of asymmetry fixed by $\mu $. Values
of $G<1$ are reflected in distributions that are barely peaked at the
corners. The polarization increases for $G=1$ and values of $G$ that are
only slightly larger than 1 produce distributions $P_\infty ^{GMG}(p)$ that
vanish for all values of $p$ except in the close neighborhood of $p=0$ and
$p=1$. Values of $G$ that are even larger places the system into the
quenching region of the phase diagram because every successful initial move
produces such a large gain that a revision of the strategy becomes extremely
unlikely; therefore the evolution of the system produces only mild changes
in the initial distribution.

\section{The BAM model}

The dynamics of the GMG does not keep track of the situations that lead to
succeses or failures. This is instead stressed in the BAM. It has been
customary to accept that agents derive utility in two situations: one when
they decide to go to the bar and $M\leq \mu $ and also when they refrain to
go to the bar and $M>\mu $.

We propose for the BAM an adaptive dynamics that is based upon the same
information as in the GMG but uses it differently in order to keep track of
the situations that lead to gains or losses i.e. by correlating them with
the action of going or not going to the bar. To this end we introduce
``points'' and ``credits''. Points are gained or lost if the option of {\em
attending} the bar is correct or not. Credits are instead gained or lost
depending whether the option of {\em not attending} the bar is correct or
not. We therefore carry a double account. If an agent chooses to attend (not
to attend) the bar and $M<\mu $ ($M>\mu $) gains a point (a credit).
Otherwise, she loses a point (a credit). When {\em either} the number of
points or the number of credits falls below some threshold, the
corresponding account of points or credits is reset to zero and the strategy
is changed in the same fashion as already mentioned for the GMG.

Note that the attendance strategy of an agent is changed by either of two
reasons. Such updating rule makes use of information that is lost with the
scoring rules used for the GMG. Suppose for instance that an agent having
initially all her accounts in zero chooses to go to the bar one day in which
$M<\mu $ and chooses not to go the next day when the attendance turns out to
be again $M<\mu $. Within the GMG rules she ends up with zero points and her
strategy is not updated. With the double account of the BAM she ends up
instead with one positive point and one {\em negative} credit and is
therefore forced to update her strategy \cite{footnote1}.

We have performed numerical simulations using this dynamics with similar
settings than those reported in the preceeding section. The results for the
BAM are compared with those obtained with the GMG dynamics in Figs.~\ref
{ejemplos1},\ref{ejemplos2}. When quenching is absent in the GMG, the
asymptotic distributions $P_\infty ^{BAM}(p)$ can barely be distinguished
from $P_\infty ^{GMG}(p)$. For this case the values of the attendance $A(t)$
and $C_s(t)$ for the BAM are shown in Fig.~\ref{ejemplos1}b). A strong
difference however occurs when $\mu >\mu _{cr}$ and the GMG gets trapped in
a quenched configuration. For the same settings the BAM converges instead to
equilibrium. The asymptotic distribution $P_\infty ^{BAM}$ shows a strong
polarization and the attendance reaches the accepted level without
difficulty. The nature of the asymptotic state has no difference with that
for $\mu <\mu _{cr}$. This can be seen in Fig.~\ref{ejemplos2}b).

A more profound diference between the two model arises when the above
analysis is repeated for $G\not =1$. While the GMG dynamics can be checked
to lead to quenched configurations for essentially all values of $G$, as
shown in Fig.\ref{Fases}, the BAM model displays no quenching independently
of the value of $G$. Quenching can however occur in the BAM but for
different initial conditions than a uniform distribution. We turn to this
point in the next section.

\section{ Analytic discussion}

In this section we discuss analytically the relevant features of the
density distribution $P(p)$ as well as the occurrence of quenching within
both the GMG and the BAM.

\subsection{The density distribution}

The shape of the distribution $P(p)$ is the combined result of the actions
of the whole ensemble of agents.  Each player adjusts her attendance
attempting to minimize her individual loss.  When equilibrium is reached
the resulting $P_{\infty}(p)$ for both GMG and BAM, concentrates the
population in the immediate neighborhood of $p\simeq 0$ and $p\simeq 1,$
plus an almost vanishing contribution from intermediate values.  The ratio
of the areas below the two peaks is, essentially, equal to $\mu /(1-\mu
)$.  An intuitive guess for $P(p)$ could well be instead a sharply peaked
function centered at $p\simeq \mu$.  The difference has to be sought in
the properties of the distribution ${\cal P} (A)$ that gives the
probability of occurrence of a party of $A$ customers attending the bar.
The distribution ${\cal P} (A)$ is of course completely determined by
$P(p)$.  If $P(p) \simeq \delta(p-\mu)$ the corresponding ${\cal P} (A)$
turns out to be large for values of $A$ that differ from the optimal one,
the highly polarized equilibrium distribution that is actually generated
produces instead a ${\cal P} (A)$ that is much more concentrated near
$A=N\mu$.

Let us assume without loss of generality that all the agents distribute
themselves into $D$ different strategies $p_i;\ i=1\dots D$ according to:

\begin{equation} P(p)=\sum_i^D \frac{n_i}{N}\delta(p-p_i). \label{eq.X}
\end{equation}

\noindent where $\sum_i^D n_i=N$. With this assumption, the distribution
${\cal P}(A)$ can be written as:

\begin{equation} {\cal P}(A)= \sum_{\ell_1=0}^{n_1} \dots
\sum_{\ell_D=0}^{n_D}\prod_i^D  {\big{\lbrack}} {n_i \atopwithdelims()
\ell_i}p_i^{\ell_i}(1-p_i)^{n_i-\ell_i}{\big{\rbrack}}
\times \ \delta(A-\sum_j^D\ell_j).  \label{distrigrupo} \end{equation}

The value $(A-N\mu)^2$ measures the departure from the optimal attendance,
and therefore ${\cal C}^2=\sum_A{\cal P}(A)(A-N\mu)^2$ can be taken to
measure the square of the average cost incurred by the ensemble of agents while
adjusting their individual strategies \cite{footnote2}.

Although Eq.\ref{distrigrupo} is in general difficult to evaluate,
${\cal C}^2$ can be calculated without great difficulties.  In fact if its
definition is introduced into Eq.\ref{distrigrupo}, the summation over $A$
can be performed first, taking advantage of the $\delta(A-\sum_j^D\ell_j)$
and all the summations over the $\ell_j$ indices decouple from each other.
It results

\begin{equation} {\cal C}^2= \sum_{\ell_1=0}^{n_1} \dots
\sum_{\ell_D=0}^{n_D}\prod_i^D
{\big{\lbrack}} {n_i \atopwithdelims()
\ell_i}p_i^{\ell_i}(1-p_i)^{n_i-\ell_i}{\big{\rbrack}}
\times \ (N\mu-\sum_j^D\ell_j)^2.  \label{distrigrupo2} \end{equation}

Moreover, from Eq.(\ref{eq.X}), \  $<p>=\sum_i n_i p_i/N$ and    $<p^2>=\sum_i
n_i p_i^2/N$, thus in the limit $D \rightarrow \infty$ one gets

\begin{equation}  {\cal C}^2=
N^2(\mu-<p>)^2+N(<p>-<p^2>) \label{forsigma} \end{equation}

In Eq.\ref{forsigma} $<p>$ stands for $\int p P(p) dp$.  If $P(p)=1/N$ as
for the case of the initial conditions chosen in all the simulations in
the preceeding sections, the dispersion turns out to be ${\cal C}^2=
N^2(\mu-1/2)^2+N/6$.  We thus see that such initial condition is a
good guess for the traditional settings of the MG when $\mu \simeq 1/2$,
but is indeed very poor for the GMG when $\mu \not= 1/2$.

The equilibrium result $<p>=\mu$ that is always found, is seen to
eliminate the $O(N^2)$ terms in ${\cal C}^2$.  This is also
achieved with the guess $P(p)=\delta(p-\mu)$.  However, all $O(N)$ terms
do not cancel because ${\cal C}^2 = N\mu(1-\mu)$.  If the two
peaked equilibrium distribution is roughly approximated by
$P(p)=\frac{n_1}{N}\delta(p-p_1)+\frac{n_2}{N}\delta(p-p_2)$ one readily
sees that the $O(N^2)$ terms are eliminated when $n_1p_1+n_2p_2=\mu N$.
The $O(N)$ terms also cancel if the two peaks are $p_1=0;n_1=N (1-\mu)$
and $p_2=1; n_2=N \mu$.

From this point of view, quenched states are far from optimizing the
average aggregate cost ${\cal C}^2$ because the corresponding
$P_q(p)$ do not even cancel the $O(N^2)$ terms.  The relaxation dynamics
that tends to minimize individual losses is thus seen to also optimize a
global parameter (${\cal C}^2)$ only if equilibrium is reached.
Otherwise the relaxation stops when no agent loses.  This situation, in
general, does not corresponds to a minimal average cost.

\subsection{BAM}

In this section we discuss the occurrence of equilibrium and quenching
within the BAM.  This can only be made without any approximation if one
assumes a distribution $P(p)=\delta (p-p_o)$.  This is seen to be
appropriate except for $O(N)$ terms in the dispersion of party sizes.

Given an acceptable level of attendance $n$, if all agents have the same
attendance strategy $p_o$, the probability of occurrence of a party of {\em
less than } $n$ agents is $S(N,n,p_o)$ where

\begin{equation}
S(N,n,p_o)=\sum_{i=0}^{n-1} {{N-1  \atopwithdelims() i}
} p_o^i(1-p_o)^{N-1-i}.
\end{equation}

\noindent This is a continuous and monotonous decreasing function of $p_o$
and fulfills the following three conditions:

\begin{eqnarray}
S(N,n,0) &=&1\ \ \ \ \forall n  \label{uno} \\
S(N,n,1) &=&0\ \ \ \ \forall n  \label{cero} \\
S(N,n+m,p_o) &\geq &S(N,n,p_o)\ \ \forall p_o,m>0  \label{monotona}
\end{eqnarray}

\noindent If an agent decides to go to the bar, her account in points will
remain equilibrated when the expected rewards equals the expected fine, i.e.
when the attendance strategy $p_p$ that is common to all the agents fulfills:

\begin{equation}
S(N,n,p_p) = \frac{1}{1+G}  \label{defppuntos}
\end{equation}

\noindent If an agent decides instead {\em not} to go to the bar a similar
condition can also be written for her account in credits. This is:

\begin{equation}
S(N,n+1,p_c) = \frac{G}{1+G}  \label{defpcreditos}
\end{equation}

Given $G,N$ and $n$, the two equations (\ref{defppuntos}) and (\ref
{defpcreditos}) define the two roots $p_p$ and $p_c$ that are respectively
the probabilities with which all agents have to go if they hope to
equilibrate, respectively their accounts of points and credits.

Assume now that $G=1$. Bearing in mind Eq.(~\ref{monotona}) the two roots
fulfill $p_p<p_c$. If the agents choose to attend the bar with a probability
$p_o<p_p$, they gain points. On the other hand, if they choose to attend
with probability $p_o>p_c$ they gain credits. However, due to the fact that
$p_p<p_c$ if they gain points they lose credits and viceversa. In either
case all agents will continue to update their strategies, quenching is not
possible and equilibrium is eventually reached.

For a $\delta $-type distribution of strategies the only chance to have a
quenched phase using the BAM dynamics is to find a situation in which
$p_c\leq p_p$.  For finite $N$ this can only happen if $G>1$ (see
Fig.~\ref {sumas}).  From the above two equations it is easy to see that
the root $p_c$ becomes smaller for increasing values of $G$ and the
opposite happens for $p_p$.  A value $G_{crit}^{BAM}$ can eventually be
found in which both roots fulfill $p_p=p_c\equiv p_{crit}$.  For any
$G>G_{crit}^{BAM}$ the regions for winning points and credits start to
overlap and quenching occurs {\em within a narrow overlapping interval}
$(p_c,p_p)$.  In Fig.~\ref{QUENCHBAM} we show the two roots $p_p$ and
$p_c$ as a function of $G$.  It further turns out that for finite values
of $N$ the value of $G_{crit}^{BAM_{}}$is nearly independent of $\mu $.

The evolution of a $\delta $-type distribution centered at a value of $p_o$
depends upon the choice of $G$ and $\mu $. If $G>G_{crit}^{BAM}$, and $p_o$
is inside the quenching interval $(p_c,p_p)$ the distribution will not
change and the attendance $M$ will remain fixed at the value chosen for the
initial conditions. This is so because because all agents win both points
and credits. If $p_o$ lies outside of the quenching region, agents are
forced to update their strategies either because they lose point or because
they lose credits. The average attendance $M$ will approach the value $\mu$.
Note that the average attendance and the average of individual strategies
are completely equivalent i.e. $M=\int pP(p)dp$ . Therefore when the average
attendance approaches the quenching interval $(p_c,p_p)$ , the evolution
will progressively come to a stop. For values of $G$ up to $G\sim 5G_{crit}$
the quenching interval $(p_c,p_p)$ is always small ($\mid p_p-p_c\mid \sim
.1 $) and the convergence to a quenched phase is in general very slow. As
the mean of the distribution approaches the quenching interval, a smaller
fraction of agents have to update their strategies and therefore the
fraction $C_s(t)$ tends slowly to zero approaching very nearly a power law,
for a considerable number ($\sim 10^6$) of iterations. Since $G_{crit}^{BAM}$
is almost independent of $\mu $ this behaviour does not change with its
particular value. In the cases in which $p_o\notin $ $(p_c,p_p)$ the
corresponding (asymmetric) distributions $P_q^{BAM}(p)$ associated to
quenched configurations look similar to those associated to equilibrium
because involve agents with all possible attendance strategies.

\subsection{GMG}

A similar analysis as the one performed for the BAM can be made for the GMG.
If the system is prepared with a distribution $P_0(p)=\delta(p-p_o)$ the
{\em single} account of points remains equilibrated provided that the
probability of winning $P_w(N,n,p)$ fulfills:

\begin{equation}
P_w(N,n,p_o)=p_oS(N,n,p_o)+(1-p_o)(1-S(N,n+1,p_o))= \frac{1}{1+G}
\label{PganarGMG}
\end{equation}

In Fig.~\ref{sumas} we show an example of the functions $S(N,n,p)$,
$S(N,n+1,p)$ and $P _w(N,n,p)$ for typical values of $N$ and $n$, and
$n>N/2$.

The situation for the GMG is completely different than for the BAM.  The
function $P _w(N,n,p)$ has one maximum in the interval $0 \leq p \leq 1$,
and to a high degree of accuracy displays the symmetry
$P_w(N,n,p)=P_w(N,N-n,1-p)$.  Symmetry however is not strict.  To see this
note that if $n=N$, going to the bar is always a winning option no matter
the option of the rest of the agents.  If $n=0$, the action of not going
is not enough to win, because that depends upon the choice made by all the
other agents.

A critical value $G_{cr}^{GMG}$ and a critical probability of attendance
$p_{crit}^{GMG}$ can be defined for the GMG as the $p$ value that
corresponds to the maximum of $P _w(N,n,p)$ given by Eq.(~\ref{PganarGMG})
through

\begin{eqnarray}
P_w(p_{crit}^{GMG}) &=& P_w^{max} = {\mbox{max}_p}{P _w(N,n,p)} \\
G_{cr}^{GMG} & = & \frac{1}{P_w^{max}} -1  \label{griticoGMG}
\end{eqnarray}

\noindent For a given value of $n$ and $N$ quenching is produced if
$p=p_{cr} $ and $G\geq G_{cr}$. the values of $G_{cr}$ for the GMG are
completely different from those of the BAM. While in the latter there exists
a common value of $G_{cr}$ independent of $\mu $ this is not the case for
the GMG (see Fig.~\ref{Fases}).

Above $G_{cr}$ and initial condition $P_0(p)=\delta (p-p_1)$, the quenching
condition is fulfilled for any $p_1$ in the interval $p_{<}<p<p_{>}$ that
are the two roots of Eq.(~\ref{PganarGMG}). Any system prepared with $p_1$
within that region will rapidly be quenched. Assume for the sake of
concretness that we choose $\mu >0.5.$ The system evolves quite differently
if $p_o<p_{<}$ or $p_o>p_{>}$. In the first case the whole population of
agents is forced to update their strategies in order to change the
individual $p$'s in order to surpass the value $p_{<}$. Once this happens,
updating progressively comes to a stops as the average attendance gets into
the interval $(p_{<},p_{>})$ and the system gets quenched with a value of $M$
below the accepted fraction $\mu $. The fact that $P_w(N,n,p)$ is such that
$p_{>}$ is very nearly similar to $\mu $ causes that for $p_1>p_{>}$ the
initial condition of a $\delta -$type distribution changes into $P_\infty
^{GMG}(p)$ with a mean value that is very close to $\mu $ that can not be
distinguished from an equilibrium distribution. In addition, updating of
strategies continues to take place continuously. Examples are shown in Fig.
\ref{exGMGdelta}.

\subsection{Finite-size effects}

An estimate of the finite size effects of the models on the values of
$G_{cr} $ and $p_{cr}$ can easily be made for the usual settings of the
MG, by assuming a $\delta -$type distribution as an initial condition.
For $\mu =1/2 $, and from Eqs.(\ref{PganarGMG}) and (\ref{griticoGMG}) it
follows immediately that $p_{crit}=1/2\ \ \forall N$ and the value of
$G_{crit}^{GMG} $ can directly be estimated from the value of
$P_w^{GMG}(p=1/2)$ using Eq.  (\ref{PganarGMG}) and standard asymptotic
expansions.  It turns out:i

\begin{equation}
G_{cr}^{GMG}=\frac {\surd[2\pi N]+2}{\surd[2\pi N]-2} \label
{finitesize}
\end{equation}

If this result remains valid for a uniform initial distribution, the usual
settings for the MG ($G=1$ and $\mu =1/2$) correspond to a critically
quenched system.

In Fig \ref{scalingBAMGMG}a) are shown $G_{crit}^{GMG}(\mu =1/2)$ and
$G_{crit}^{BAM}(\mu =1/2)$ as a function of $1/N$. It is seen that
$G_{crit}^{GMG}\simeq G_{crit}^{BAM}$ as $N\rightarrow \infty$.The
corresponding values for $\mu \not =1/2$ can be related to those shown in
Fig.~\ref{scalingBAMGMG}a). In the first place one should notice that the
value of $G_{crit}^{BAM}$ is independent of $\mu $. On the other hand, the
curves of $G_{crit}^{GMG}(\mu \not =1/2)$ can be reduced to the one shown
through $G_{crit}^{GMG}(\mu )=(a\mu +b)G_{crit}^{GMG}(\mu =1/2)$ to a very
good degree of approximation ($a$ and $b\sim $ constants, up to terms
$O(1/N) $).

In what refers to the values of $p_{crit}$ one can check (Fig.~\ref
{scalingBAMGMG}b)) that $p_{crit}^{BAM}$ and $p_{crit}^{GMG}$ approach each
other and tend to the common limiting value $p_{crit}^\infty =\mu $, as
expected from a na{\"\i }ve estimate of the probability of attendance. One
can also conjecture that $G_{crit}^{BAM}(\infty )=1$.

\section{Annealing}

The situations described hitherto indicate that outside a very restricted
region of $\mu $ and for different initial conditions, the dynamics of the
GMG rapidly approaches quenching.  The many agent system stops evolving,
the attendance remains fixed at a value that is far from the tolerated
fraction $\mu $ and all the other internal parameters remain close to the
initial conditions.  Opposed to this, the BAM approaches always a
distribution $P_\infty ^{BAM}(p)$ that corresponds to what one expects of
thermodynamics equilibrium:  the attendance fluctuates around the expected
fraction $\mu $, agents continue to update their individual strategies and
$P_\infty ^{BAM}(p) $ remains stationary.

There is a way to regain equilibrium within the framework of the GMG through
an iterative procedure that can be regarded as an annealing protocol. The
key point is to realize that quenching is produced by the memory stored in
the points gained by the agents. In fact the quenched state remains
unchanged because on the average agents continue to gain points. To correct
the quenched phase, a procedure can be implemented that consists in
obliterating periodically the memory of the system. This is donne removing
all the points gained by all the agents in each iteration. In each annealing
episode the system is allowed to evolve freely during a maximum of $n_s$
steps; at the end, several parameters are calculated: the fraction $\varphi
_5$ of agents that have gained not less that 5 points, the relative
attendance $M,$ the fraction of histories $f_h$ where the system does not
reach a quenched configuration in the allowed $n_s$ steps, and the fraction
$f_q=n_q/n_s$, where $n_q$ is the actual number of evolution steps that
are needed to quench the system.

The results obtained are shown in Fig.\ref{annealingGMG} for two values of
$G $ and $\mu =850/1001$.  At the end of the first episode the
distribution $P_q$ (Fig.\ref{annealingGMG}a)) that is found is entirely
similar to the humped distribution shown in Fig\ref{ejemplos2}, and
$\varphi _5=1$, thus corresponding to a quenched phase.  As agents are
repeatedly deprived of points some of them are increasingly forced to
update their strategies.  For $G=1$ (Fig.\ref{annealingGMG}b)) and during
the first 30 annealing episodes very few steps are required in order to
have $\varphi _5=1$; after $\sim $ 35 annealing episodes a clear
transition takes place that can be interpreted as a gradual melting of the
system:  the fraction $\varphi _5$ drops while $f_{h\text{ }}$and $f_q$
grow abruptly, almost reaching their maximum possible values.  For $G=1.1$
(Fig.\ref{annealingGMG}c)) this transition is not present as can be seen
from the fact that $\varphi _5$ is constantly equal to one, while $f_h=0$.
It is most remarkable, however, that the relative attendance $M$ still
reaches the value $\mu $, as for $G=1$.  In other words, the annealing
protocol is seen to be able to drive the system towards $M\simeq \mu $
with or without a change of the phase of the system.

\section{Conclusions}

In the present work we have discussed the self organization properties of
two multi agent models, the BAM and the GMG. We have shown the relevance of
two control parameters. One is the fraction of the population of agents $\mu
$ that has to be joined in order to make the correct decision. The other is
$G$ that represents the gain produced by a correct decision, measured in
units of the fine payed for making the wrong one. For the sake of
concreteness we have borrowed the framework of the BAM in which the decision
that has to be made is to go or not to a bar. The decision is the correct
one if the fraction of the population that has decided to go does not exceed
$\mu $.

The system under consideration lacks any interaction or internal correlation
among the agents. Each agent feels the presence of the rest only through the
value of an aggregate parameter that is the total attendance and rewards or
penalties of individual actions are decided with respect to its particular
value in each moment. Otherwise, agents act independently from each other.
With these rules, the many agent system adapts and always reaches stable
asymptotic configurations. These are, however, of quite different nature
depending upon the values of the control parameters, of the initial
conditions and on the information that is used to guide the individual
actions.

The usual framework of the minority game, if generalized to take into
account arbitrary values of $\mu $, gives rise either to states of
equilibrium or to quenched configurations. The former have a density
distribution of individual strategies that only depends upon the value of $G$
and $\mu $ and remains stationary while a fraction of the population $C_s$
continuously updates their strategies. Quenched states have instead
frequency distributions that may be very different depending upon the
initial conditions, while $C_s$ drops exponentially to zero in all cases.

A phase diagram for the GMG can be established in the $\mu -G$ plane.
Quenching occurs outside a region limited by a critical line $\mu
_{cr}(G)$.  This border line has been obtained numerically for uniform
initial conditions.  If a $\delta -$ type distribution is assumed analytic
expressions can be obtained.  In both cases finite size effects are found.
A rigorous estimate can be made for an infinite system in this later case.
It is found that if $N\rightarrow \infty $ the point $(G=1,\mu =1/2)$
belongs to the critical line.  A numerical estimate of finite size effects
for a uniform initial distribution is also in agreement with this result.
It is interesting to note that these are the usual settings for the
minority game widely used in the literature.

The shape of the distribution $P(p)$ is the combined result of the actions
of many `independent' agents. In spite of such independence we have
proven that the equilibrium distribution optimizes the statistical
occurrence of the most convenient coordinated state or, equivalently,
minimizes the average aggregate cost of the ensemble of agents.

A quenched state is reached when the individual updating of strategies
stops. This should however not be considered to correspond to a ``frozen''
configuration. If this where the case, an internal ordered state would be
found bearing no relationship with the initial conditions. Contrary to this,
the quenched state is far from being unique, the ``glassy'' structure of the
quenched phase lacks any preferred internal ordering. This should be
regarded as the signature of the absence of interactions among the agents
and has to be attributed purely to dynamic effects. While equilibrium is
associated to the optimization of the global magnitude, this is no longer
true for a quenched configuration.

Reaching a quenched phase is a direct consequence of the evolution dynamics
contained in the rules for updating the individual strategies. This effect
is evident when the GMG is compared to the dynamics arising from the BAM
with mixed strategies, as we have presented here. Contrary to the GMG, when
starting from a uniform distribution the BAM always reaches a state of
equilibrium, independently of the values of $\mu $ and $G$. The two models
differ in the use of the available information. While the GMG dynamics only
keeps record of good and wrong decisions, the seemingly minor modifications
implied in the BAM rules, allows to keep also track of the {\em reasons} for
such successes and failures. The system thus keeps a better ``memory'' of
the previous steps in the dynamics. The consequence of this is that the
system always self organizes in such a way that $M\rightarrow \mu $.

The many agent system that has been presented here has several features
that are reminiscent of the physics of many body systems.  In the first
place one should note that a quenched phase can be induced for critical
values of $\mu $ and $G$.  A sort of order parameter can be established
with the relative attendance $M(t)$.  Two regions can be individualized in
parameter space, one in which $ M(t)\rightarrow \mu $ and the other in
which $M(t)\rightarrow $ $M_\infty \not =\mu $. For critical values of
$\mu $ and $G$, the system undergoes an internal transformation reflected
by a peak in the dispersion $\sigma_A$ that reminds the behavior of the
susceptibility at a second order phase transition.

Although the model does not contain any two body interaction, within this
analogy ``particles'' accomodate themselves following an
ocupational constraint (the maximum allowed attendance) very much in the
same fashion as in many fermion system.  To support this viewpoint, one
could note that very few ``particles'' participate in the dynamics.  The
parameter $ \mu $ fixes the total attendance and therefore plays the role
of the chemical potential through an effective collective interaction.
From this point of view individual gains should be related to (the
negative of) a kind of single particle energy:  those agents that have
accumulated large gains are less likely to participate in the dynamics.

A resemblance can also be established between quenching in GMG and similar
situations found in many body systems. In fact we have described a gradual
modification of the quenched configuration through a procedure that is
entirely similar to an annealing protocol. It turns out that a quenched
state can gradually be ``melted'' into a state of equilibrium if all the
points that have been gained by the agents are periodically eliminated. The
points accumulated by the agents keep memory of the previous results and
each remotion can be thought as an exchange of  `energy' with a thermal
bath in order to reach equilibrium.

The concept of temperature has been introduced in the MG in several
contexts.  In all cases it is assimilated to random fluctuations.  In this
work we have not attempted  to introduce the temperature. It is worthwhile
to mention, however, that in fact {\em there is} an individual source of
fluctuations present, associated with the update of $p_i$ described in
Section \ref{sec:GMG}. We recall that every time a new value of $p_i$ has
to be chosen, it is randomly selected from an interval of width $\delta p$
around it. Hence, this fluctuation acts as a source of fixed temperature,
$T_0$ say, and can be recognized as the final reason that explains why the
distribution function $P(p)$ does not simply converges to the sum of several
$\delta$'s. Hence, our work can be considered to describe the relaxation
to equilibrium at that fixed temperature $T=0$, except when the annealing
protocol is applied. This framework could be extended to a situation at
$T\not =0$ by introducing fluctuations allowing for instance individual
agents to change probabilistically their assistance strategies.

The BAM has been worked out to show how internal coordination of markets
or economic systems can derive from uncoordinated individual actions that
are inspired by a common belief.  It is a clear example in which an
ensemble of agents with limited information about somebody else's
decisions can nevertheless give rise to a self organized pattern within a
scheme referred to as ``bounded rationality''. The relaxation process
reminds what in economics is known as ``tatonnement".  Opposed to this,
the ``rational expectations" solution to the BAM or the GMG would be a
distribution $P(p)=\delta(p-\mu)$ because it describes the only common
action that is consistent with the general {\em knowledge} of all agents
about everybody else's decisions concerning the acceptable attendance.  It
is interesting to note that this solution is in general far from
equilibrium and hence is neither stationary nor optimal.

The Minority Game, regarded as a simplified version of the BAM has also
been repeatedly presented as a schematic framework with similar interest
for economics.  However the occurrence of quenched asymptotic
configurations in a straightforward generalization of the MG puts some
limitations to the aplicability of the model to stylized economic
situations.  Quenched configurations are trivial examples of equilibria
because the adaptation dynamic has stopped, but they are neither Nash
equilibria nor Pareto optimal.  In fact, if $M\not =\mu $ many individual
situations can be improved without harming the situation of any other
agent. It is intersting to note that such configurations are produced by
the limited use that agents make of the available information.  GMG can
therefore safely be used if quenching is avoided or special situations are
considered in which this is not particularly relevant.  Otherwise, one
should bear in mind that to simulate market environments it is mandatory
to assume that the agents must make full use of all the available
information.

E.B. has been partially supported by CONICET of Argentina, PICT-PMT0051;H.C.
and R.P. were partially supported by EC grant ARG/b7-3011/94/27, Contract
931005 AR.

\pagebreak

\begin{figure}[tbp]
\caption{GMG and BAM without quenching.  In this case, $N=1001$, $G=1$ and
$\mu=600/1001$.  The data is averaged over 200 histories.  In (a) we show
$P_{\infty}(p)$ for both cases; (b) illustrates the time evolution of the
reduced attendance $M(t)$ and $C_s(t)$, the fraction of moving agents.
Notice that $C_s(t)$ {\em does not vanish}}
\label{ejemplos1}
\end{figure}

\begin{figure}[tbp]
\caption{BAM and GMG with quenching, for the same values of $N$  and $G$
than in the previous figure, but $\mu=800/1001$. In (a) we show that the
shape of $P(p)$ has little changes for the BAM, while the GMG distribution
is strongly modified. Correspondingly, in (b) one can see that GMG does not
reach the allowed assistance ($M \neq \mu$) because the system stops
evolving ($C_s=0$) very rapidly. Data for GMG is averaged over 3000
histories. In the broken x-axis we  show that BAM does not quench even
after $10^6$ time steps}
\label{ejemplos2}
\end{figure}

\begin{figure}[tbp]
\caption{(a) Reduced attendance as a function of $\mu$, for $N=101$. It
can be seen that $\mu_{cr}$ is a function of $G$. Size effects are shown
in the inset. (b)  transition from a non-quenched ($\varphi _{20} \neq 1$)
to a quenched state ($\varphi_{20}=1)$}
\label{Concurrencia}
\end{figure}

\begin{figure}[tbp]
\caption{Phase diagram for the GMG in the $\mu - G$ plane.  ($\triangle$)
shows an analytic solution for a $\delta$-type initial condition, while
($\Box$ and ($\Diamond$) correspond to numerical solutions starting from a
uniform distribution. The critical border for the BAM with $\delta$-type
initial conditions would be a vertical line at $G \simeq 1,22$ for $N=100$}
\label{Fases}
\end{figure}

\begin{figure}[tbp]
\caption{Change of $P(p)$ with $G$, for $\mu=0.6$}
\label{baldes}
\end{figure}

\begin{figure}[tbp]
\caption{The roots $p_p$ and $p_c$ of Eqs.(\ref{defppuntos}) and
(\ref{defpcreditos}) are exemplified for $N=20$ and $n=14$.  These values
are chosen for the sake of clarity of the figures.  Notice that as $G$
increases $p_c$ becomes smaller and $p_p$ becomes larger}
\label{sumas}
\end{figure}

\begin{figure}[tbp]
\caption{Quenching for the BAM for $\delta$-type initial conditions is possible
for $G>G_{cr}$. The roots $p_p$ and $p_c$ are ploted as a function of $G$ and
the indicated values of $\mu$. Both roots are equal for $G=G_{cr}$. The value
of $G_{cr}$ is largely independent of $\mu$.}
\label{QUENCHBAM}
\end{figure}

\begin{figure}[tbp]
\caption{Distributions $P(p)$ for the GMG (right axis), obtained from
three initial conditions $\delta(p-p_0)$. We also show $P _w$, (left
axis) to help to understand the evolution of the systems. The roots $p_<$
and $p_>$ correspond to the two roots $P _w=1/2$. In this case, $\mu=75/101$}
\label{exGMGdelta}
\end{figure}

\begin{figure}[tbp]
\caption{(a) $G_{crit}$ {\em vs} $ 1/N $ up to $N=1000$. Eq.\ref
{finitesize} provides an excellent fit that can not be distinguished from
numerical data for GMG, $G=1, \mu=0.5$ ; (b) $p_{cr}$ as a function of $1/N$
for the same values of $\mu$.}
\label{scalingBAMGMG}
\end{figure}

\begin{figure}[tbp]
\caption{Annealing in the GMG}
\label{annealingGMG}
\end{figure}

%
%

\begin{references}
\bibitem{footnote2} Note that ${\cal C}$ differs from $\sigma_M$
introduced previously, because the former is the dispersion at time $t$
with respect to the {\em allowed attendance} $N\mu$, while the later is
the dispersion of the {\em asymptotic} attendance with respect to its
average.
\end{references}
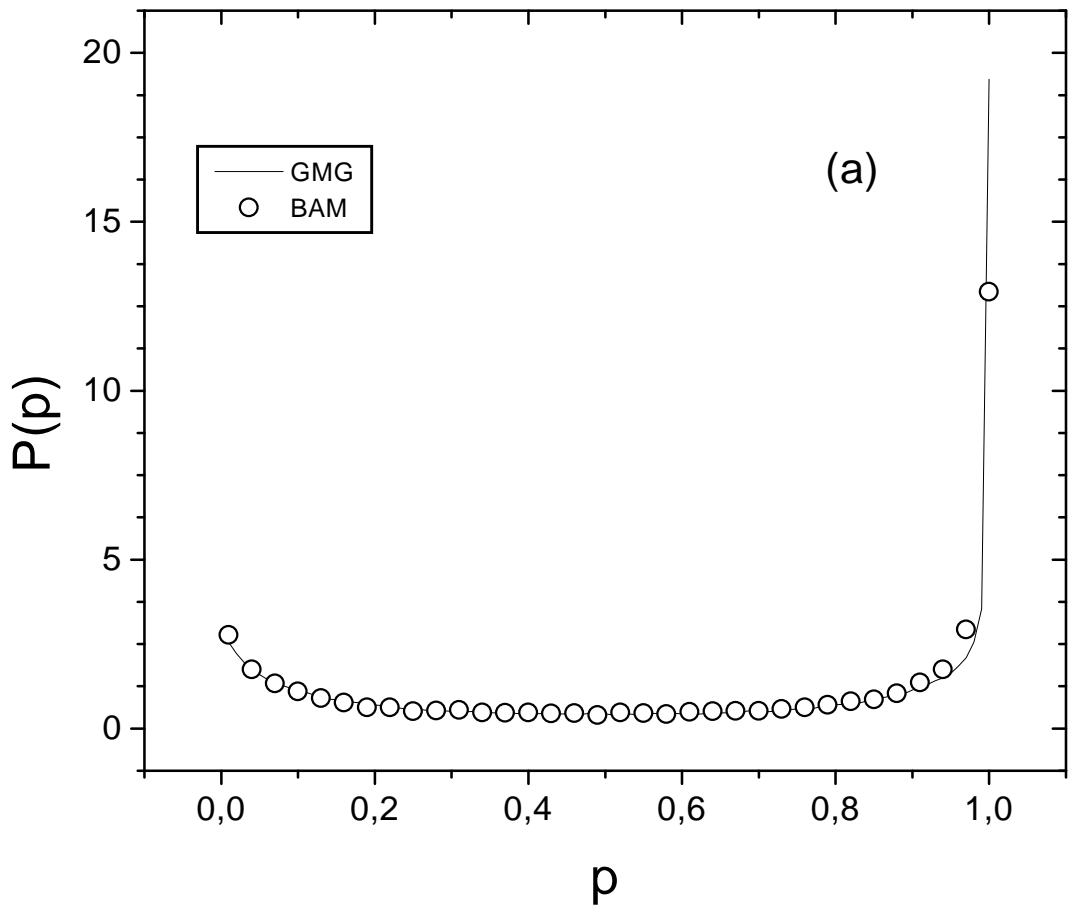
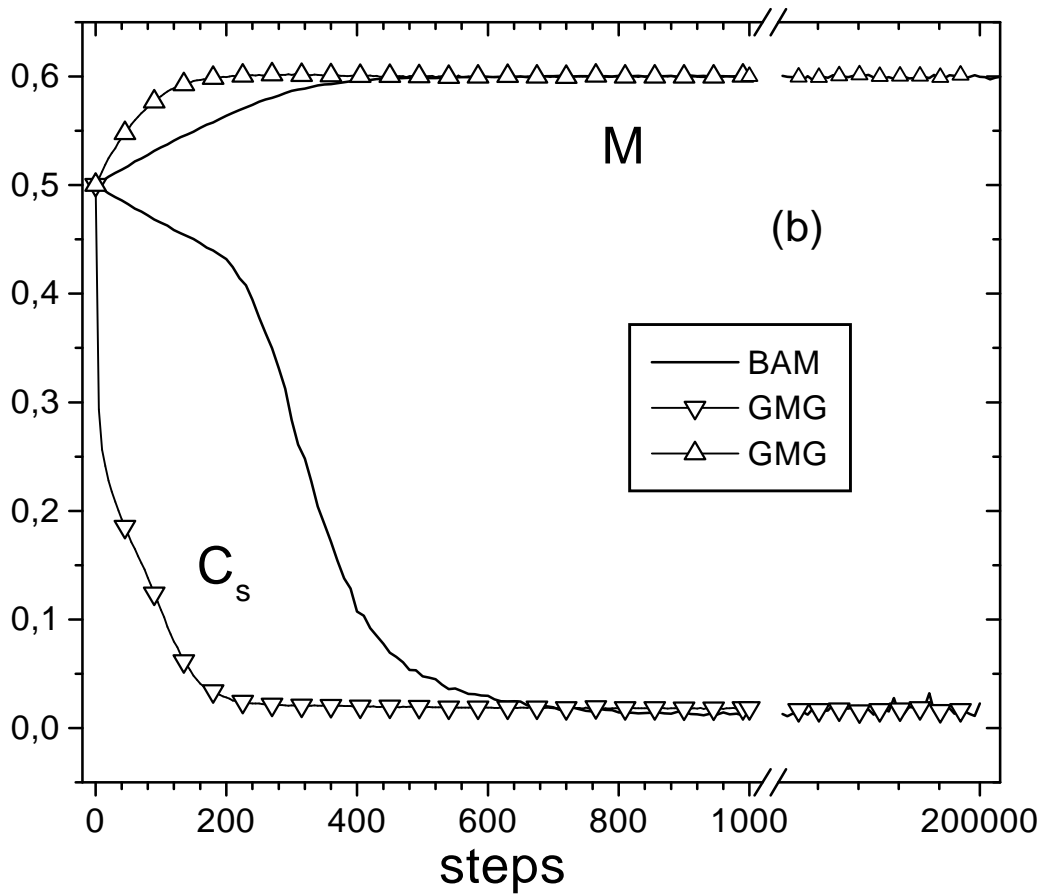
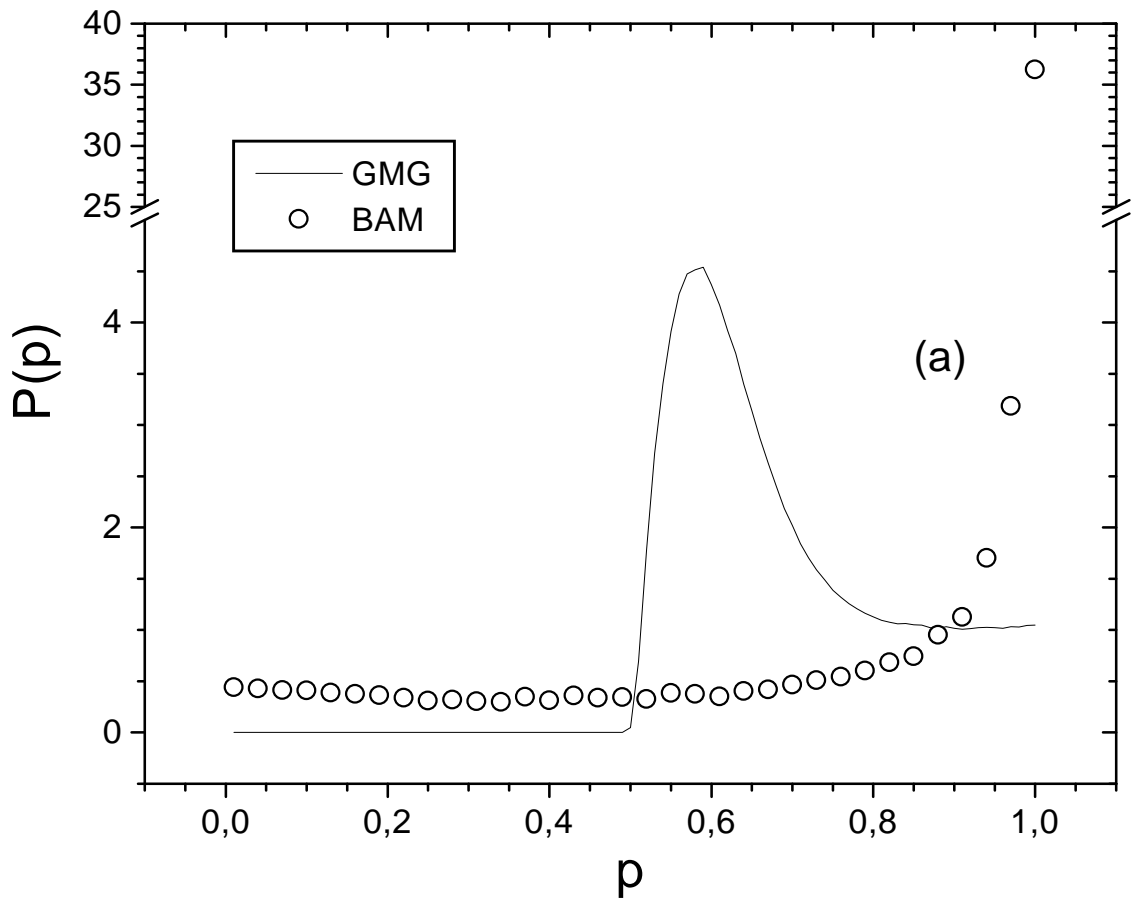
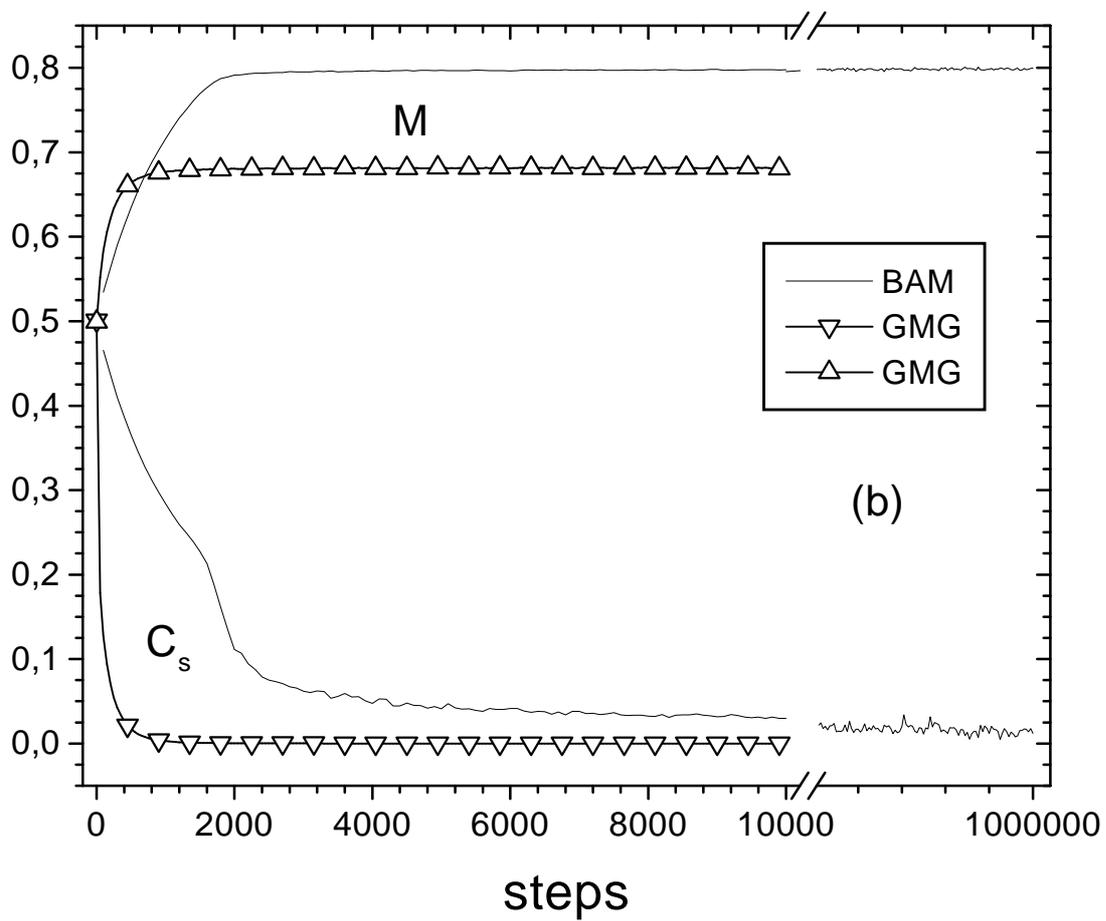
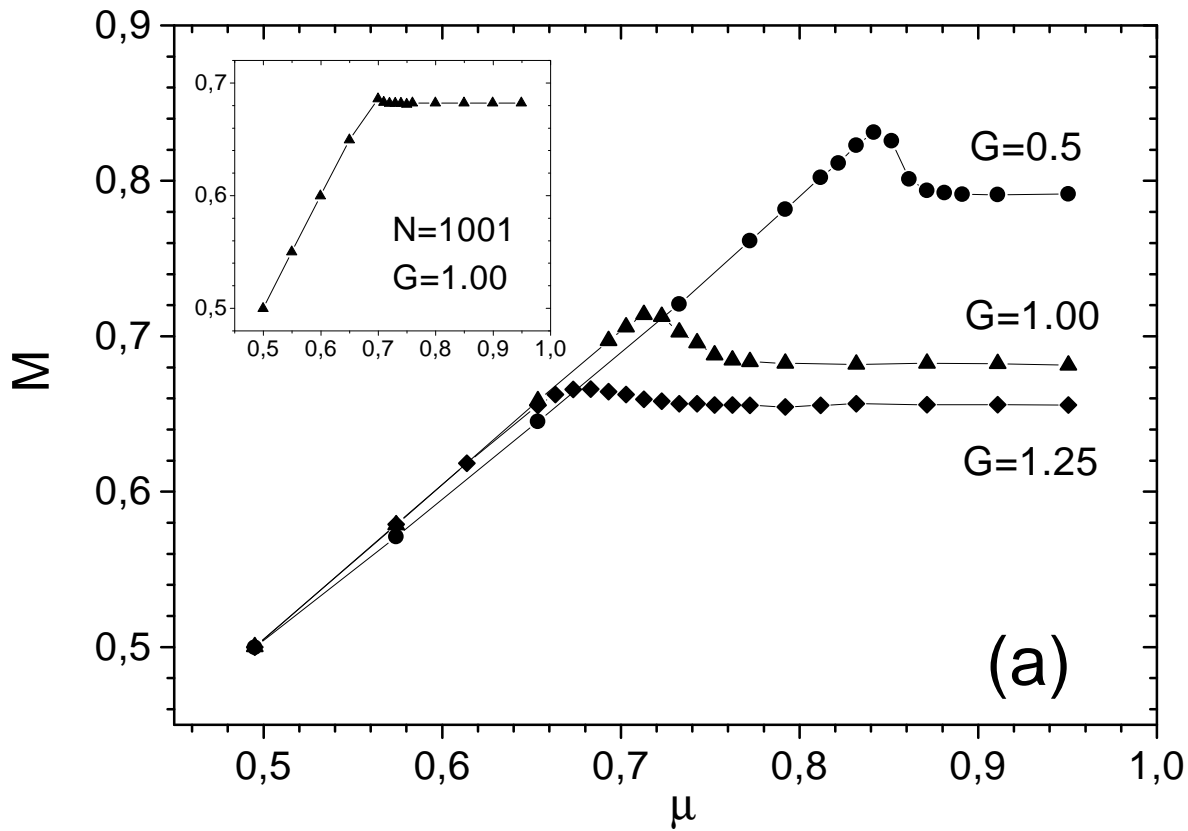
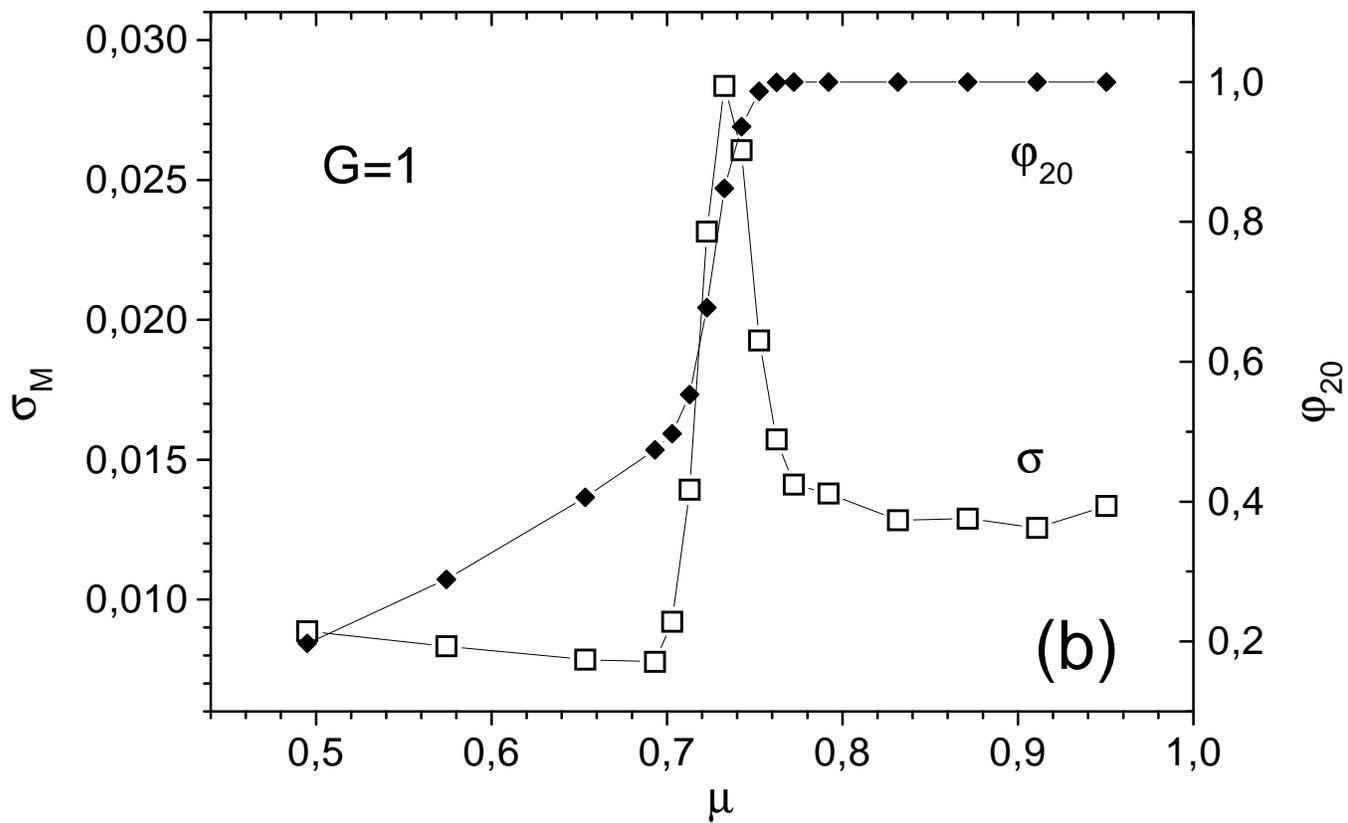
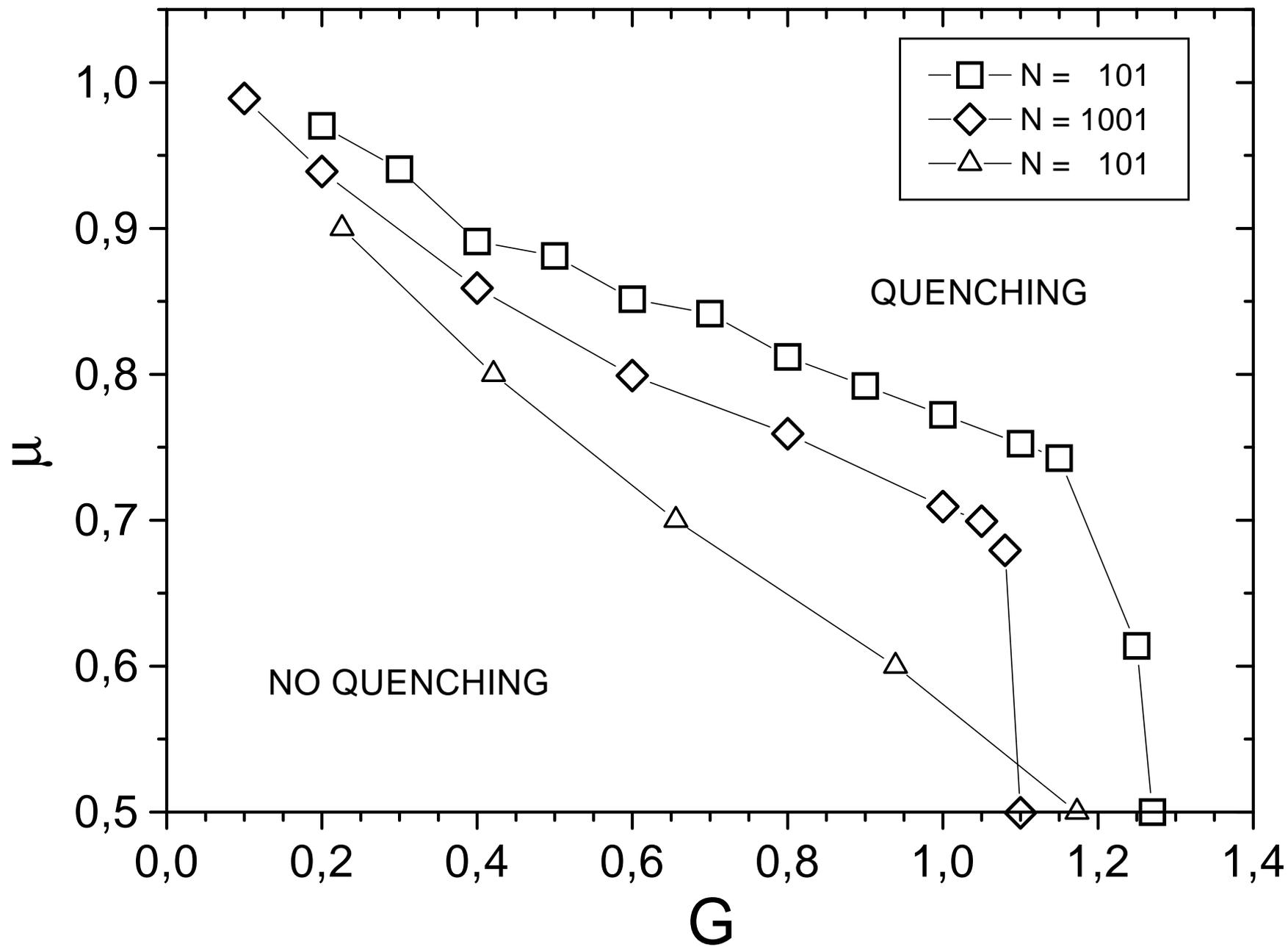
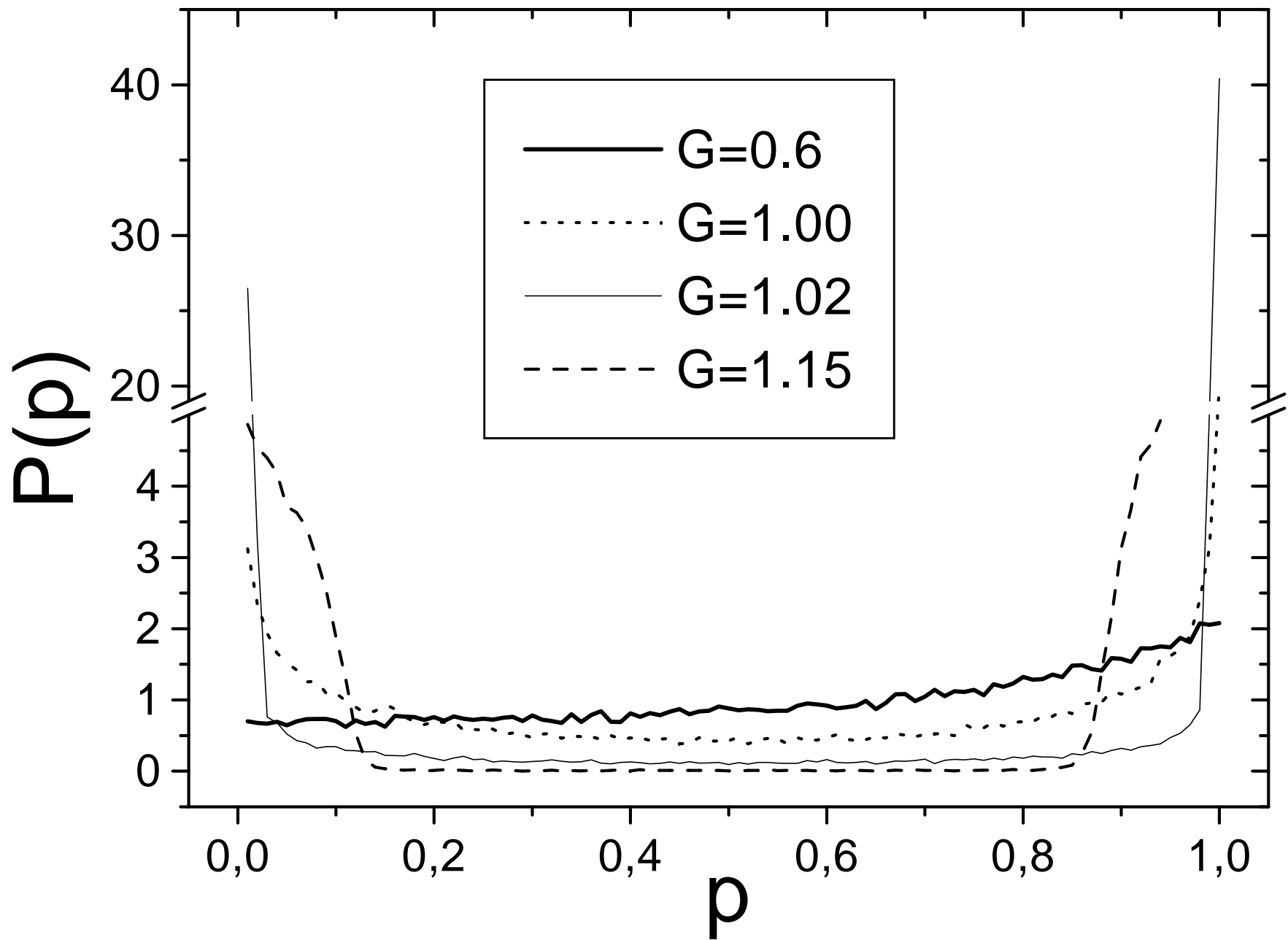
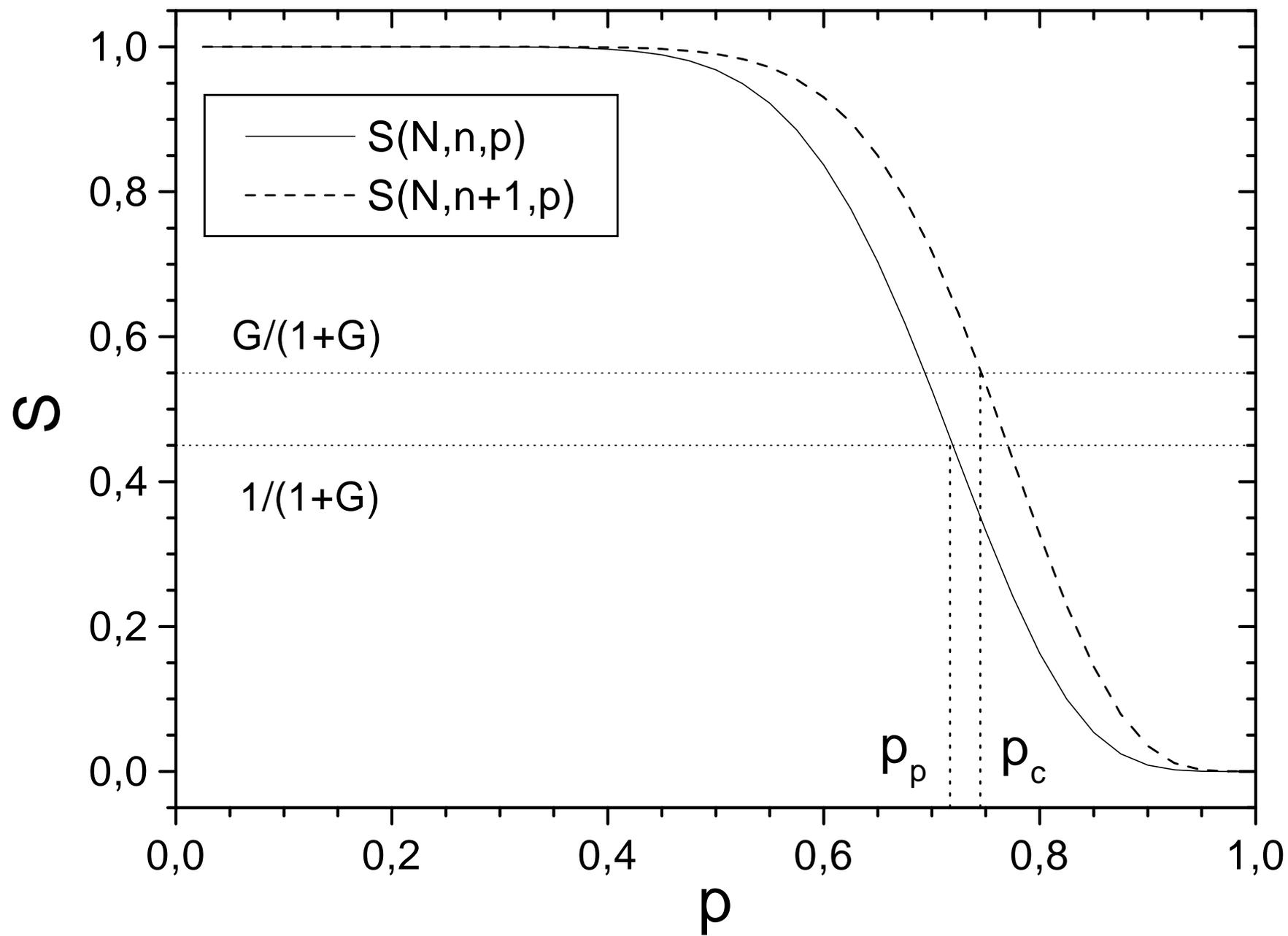
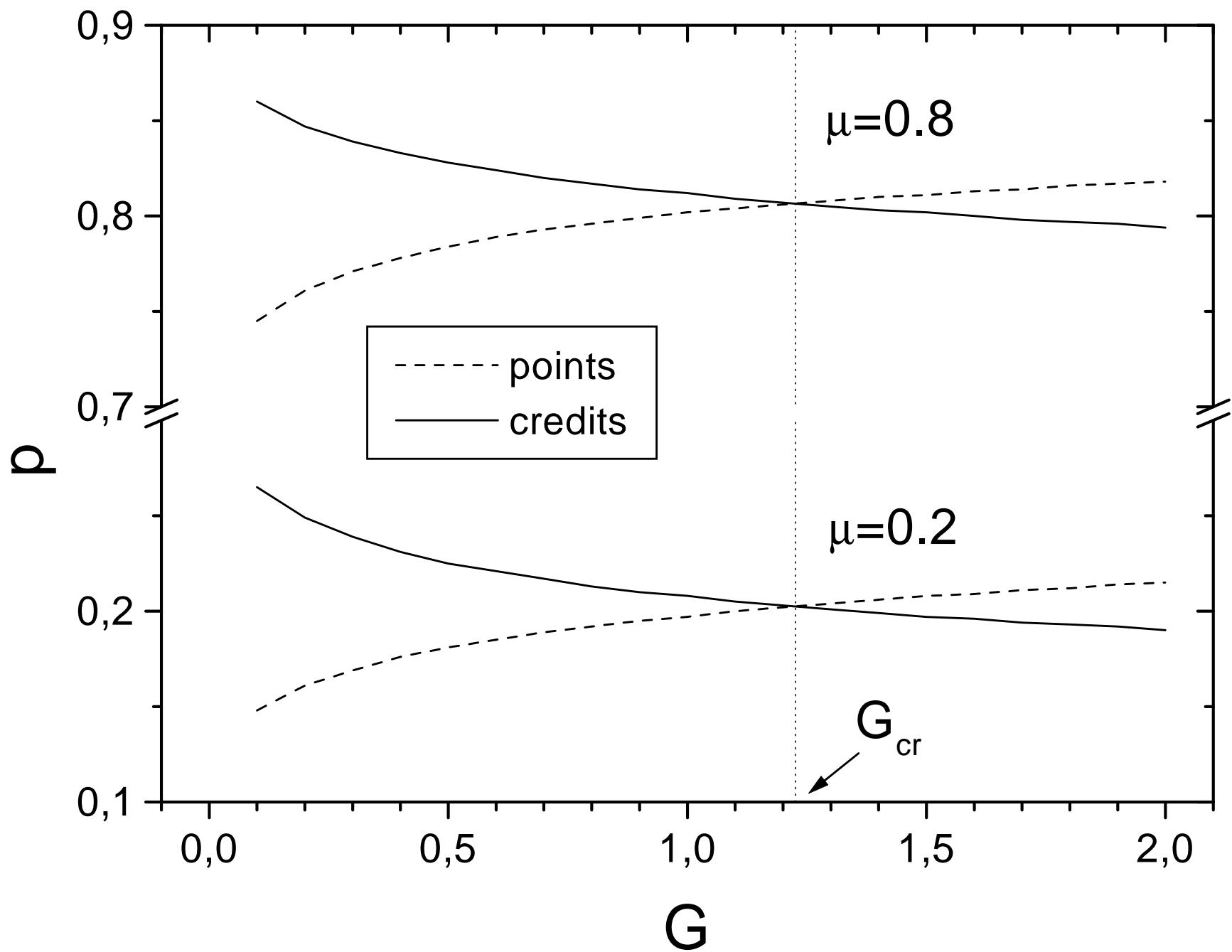
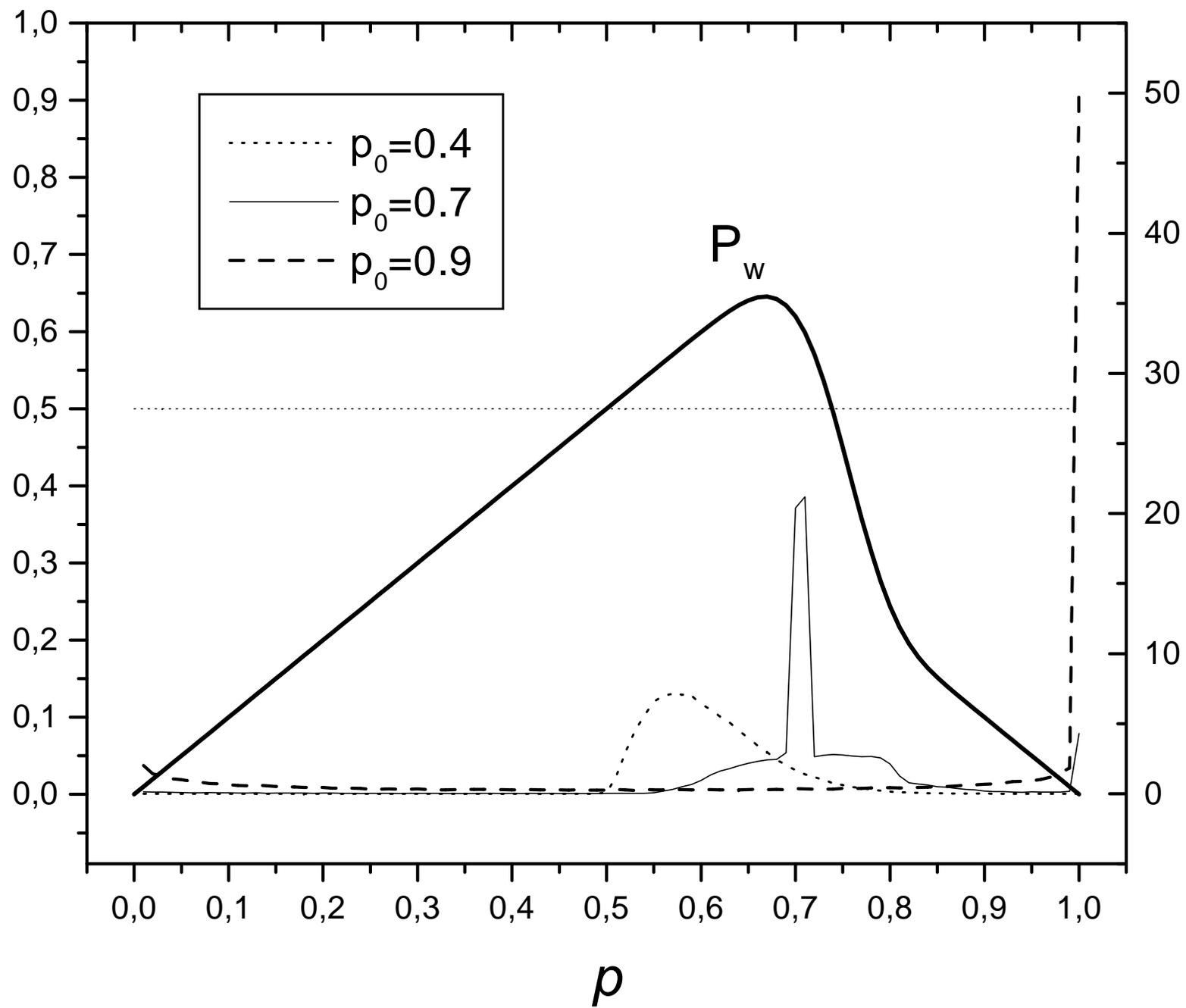
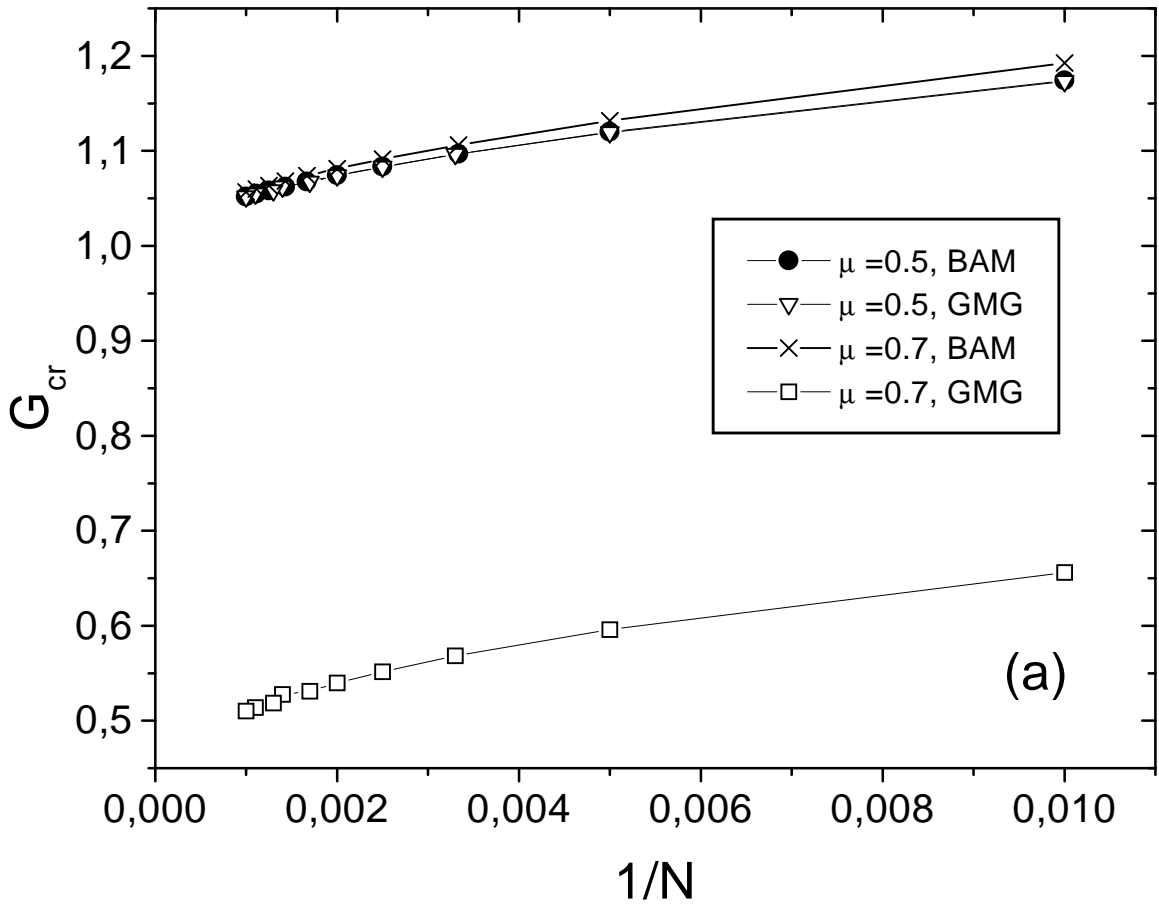
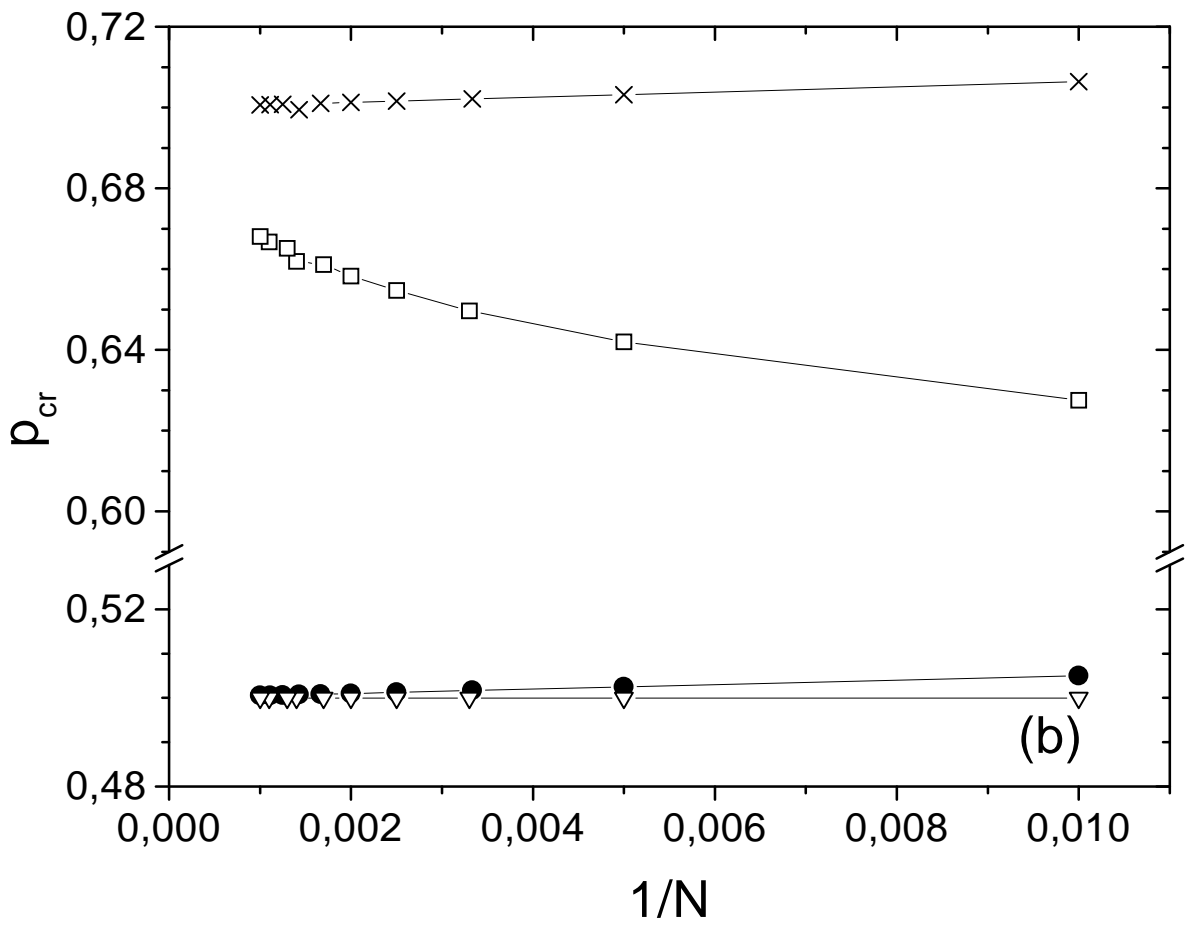
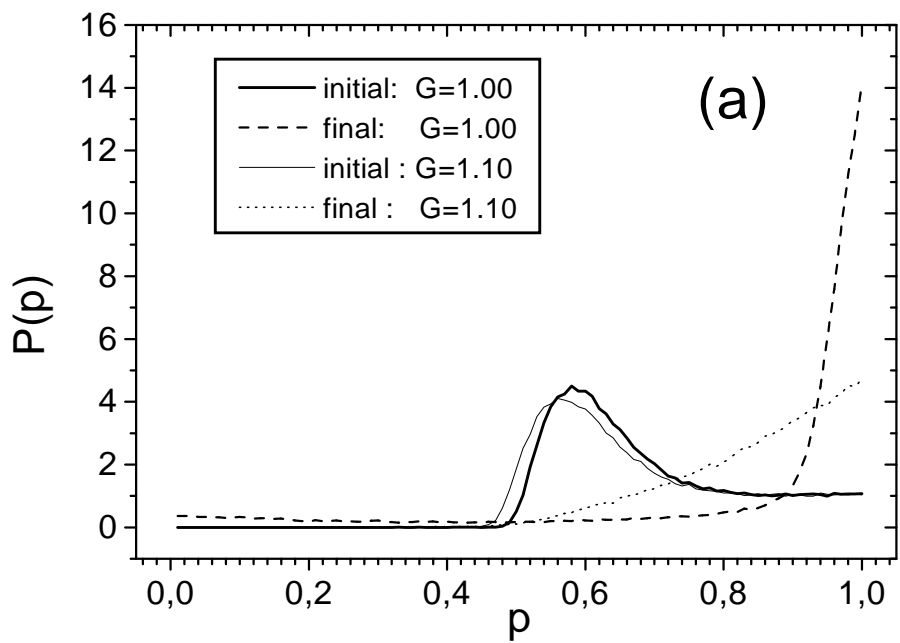
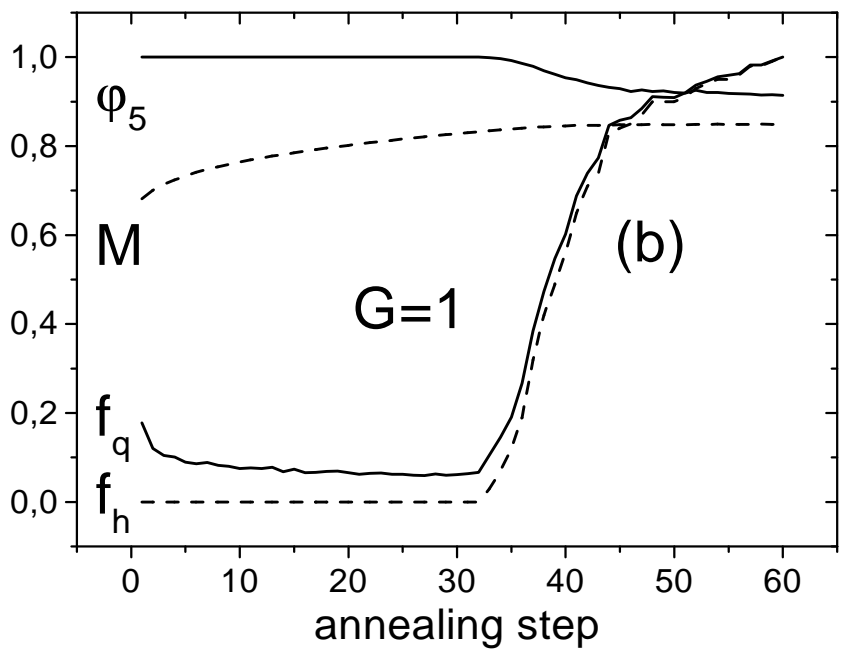
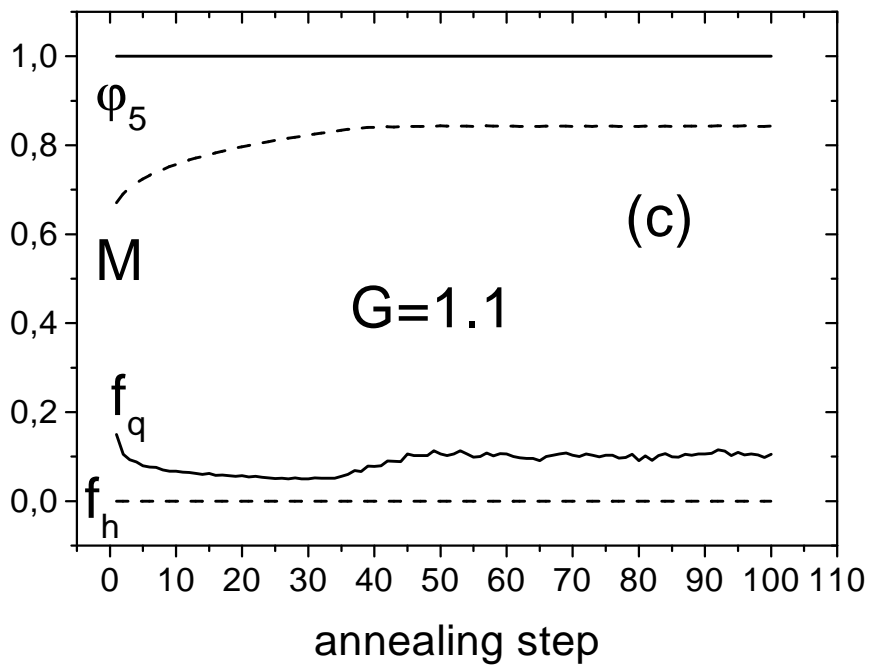
\end{document}